\documentclass[11pt,preprint2]{aastex}  %preprint   1-column, single-spaced

\slugcomment{accepted for publication in ApJ}

\shorttitle{Universal Temperature Profile}
\shortauthors{Loken et al.}

\begin{document}
 
\title{A Universal Temperature Profile for Galaxy Clusters}
 
\author{Chris Loken\altaffilmark{1}, Michael L. Norman\altaffilmark{2}, 
Erik Nelson\altaffilmark{2}, Jack Burns\altaffilmark{3},
Greg L. Bryan\altaffilmark{4},
Patrick Motl\altaffilmark{3}
}

\altaffiltext{1}{Canadian Institute for Theoretical Astrophysics, 60 St.~George St., University of Toronto, Toronto, 
ON M5S 3H8}
\altaffiltext{2}{Center for Astrophysics \& Space Sciences, University of 
California, San Diego, CA}
\altaffiltext{3}{Department of Astrophysical and Planetary Sciences, 
University of Colorado, Boulder, CO 80309}
\altaffiltext{4}{University of Oxford, Astrophysics, Keble Road, Oxford, OX1 3RH}

\begin{abstract} 
We investigate the predicted present-day temperature profiles of the hot, X-ray 
emitting gas in galaxy clusters for two cosmological models - a current 
best-guess  $\Lambda$CDM model and  standard cold dark matter (SCDM). 
Our numerically-simulated ``catalogs" of clusters are
derived from  high-resolution (15 h$^{-1}$kpc) simulations which make use of 
a sophisticated, Eulerian-based, Adaptive 
Mesh-Refinement (AMR) code that faithfully captures the shocks which are 
essential for correctly modelling cluster temperatures.
We show that the temperature structure on Mpc-scales is
highly complex and non-isothermal. However, the temperature profiles of
the simulated $\Lambda$CDM and SCDM clusters are remarkably similar and
drop-off as $T \propto (1+r/a_x)^{-\delta}$ where $a_x \sim r_{vir}/1.5$
and $\delta \sim 1.6$. This decrease is in good
agreement with the observational results of Markevitch et al.~(1998)
but diverges, primarily in the innermost regions, from their fit
which assumes a  polytropic equation of state. Our result is also in good
agreement with a recent sample of clusters observed by {\it Beppo}SAX though
there is some indication of missing physics at small radii ($r<0.2 \; r_{vir}$).
We discuss the interpretation of our results and make predictions for
new x-ray
observations that will extend to larger radii than previously possible.  
Finally, we show that, for $r>0.2 \; r_{vir}$, our universal temperature profile is 
consistent  with our most recent 
simulations which include both
radiative cooling and supernovae feedback.

\end{abstract}


\section{Introduction}

X-ray based cluster mass determinations generally assume that
morphologically-symmetric (i.e. non-merging) clusters are isothermal.
However, this accepted wisdom has recently been challenged by
Markevitch et al.~(1998; hereafter M98) who found evidence for decreasing 
temperature profiles in a sample of 
nearby hot clusters ($>3.5$ keV) observed with
ASCA. A subsample of 17 regular/symmetric clusters displayed
remarkably similar temperature profiles
(when normalized and scaled by the virial radius) consistent with
$T\propto [1+(r/r_c)^2]^{-3 \beta (\gamma-1)/2}$ where 
$\gamma=1.24^{+.20}_{-.12}$ and $\beta=2/3$. The typical decrease is 
therefore a factor
of $\sim$2 in going from 1 to 6 core radii (or .09 to 0.5 virial radii).
This result remains controversial as three subsequent studies of large 
samples of clusters concluded that
the majority of cluster temperature profiles show little, or no,
decrease with radius (Irwin, Bregman, \& Evrard 1999; White 2000; Irwin
\& Bregman 2000). Most recently, De Grandi \& Molendi (2002) have presented
a composite temperature profile based on {\it Beppo}SAX data which exhibits an
isothermal core and then decreases quickly. Here we present results
derived from recent high-resolution numerical simulations and show
that there appears to be a universal temperature profile which
declines significantly even within half a virial radius of the
cluster center. We also show that these simulated profiles
are consistent with the most recent cluster observations except in the very innermost
regions of a cluster ($r<0.2 \; r_{vir}$).

\section{Simulations}

The simulated clusters used in this work are part of an ongoing project 
to create extensive numerical
``catalogs'' of clusters. This catalog has 
already been used to investigate the effect of large-scale structure on
weak gravitational lensing mass estimates (Metzler et al.~1999, 2001) and
the role of ``temperature bias'' in Sunyaev-Zeldovich determinations
of the Hubble constant (Lin, Norman, \& Bryan 2001).
The cluster simulation data are available on-line at the 
Simulated Cluster Archive: \url{http://sca.ncsa.uiuc.edu} (Norman et al.~1999).

The cosmological
model adopted for the bulk of the simulations is a current ``best-guess''
flat universe with a cosmological constant ($\Lambda$CDM) and
parameters: $\Omega_0 = 0.3, \; \Omega_b = 0.026, \; 
\Omega_\Lambda=0.7, \; h=0.7$ and $\sigma_8 = 0.928$. The 25 most-massive 
clusters in the 256 h$^{-1}$ Mpc volume have been
fully analyzed and comprise the $\Lambda$CDM sample used in this work. 
For comparison, a smaller sample of 10 SCDM clusters has been simulated 
in the same volume (and with the same random-seed for the initial fluctuations)
with parameters: $\Omega_0 = 1, \; \Omega_b = 0.076, \;, \; h=0.5$ 
and $\sigma_8 = 0.6$

An initial low-resolution simulation was used to
identify clusters in the  256 h$^{-1}$ Mpc volume and then 
high-resolution simulations were performed which evolved the entire 
volume but adaptively refined
the region from which a specific cluster formed. 
The root, or level 0 (L0), grid covers the entire periodic domain with 
$128^3$ cells. The Lagrangian volume of a given cluster is
further refined with two levels of subgrids (L1 and L2), each with twice
the spatial resolution and eight times the mass resolution of the previous one, 
in order to refine the initial conditions
within this volume. An additional five levels of refinement
are introduced adaptively and automatically
within the L2 grid as the simulation progresses.
The clusters form from particles on the L2 grid which have a mass of 
$9\times10^{9}$h$^{-1} $M$_\odot$. The best spatial resolution is 
15.6 h$^{-1}$kpc.

The cosmological adaptive-mesh refinement (AMR) code used here is 
described in detail in Bryan (1999) and Norman \& Bryan (1999). 
Briefly, this code
solves the equations of hydrodynamics using the higher-order accurate
piecewise parabolic method (PPM; Collela \& Woodward 1984) and
employs an adaptive particle-mesh algorithm (using second-order
accurate TSC interpolation) for the dark matter particles.
The Eulerian
hydrodynamics algorithm exhibits several crucial characteristics
including excellent shock-capturing (within 1-2 zones with correct entropy
generation), and equal accuracy in high or low density regions (for the
same zone size). 
The key component of the code is its use of structured
adaptive mesh refinement (Berger \& Collela 1989) in order to enhance
resolution when and where needed.

Our code was tested as part of the Santa Barbara cluster comparison project
(Frenk et al. 1999) in which 12 groups simulated a Coma-like cluster using 
a variety of codes and resolutions
Our results (curves labeled ``bryan") are among
the highest resolution results presented in that paper (central resolution
of $7.8h^{-1}$ kpc). We agree well with the highest resolution $AP^3MSPH$ results except 
that our central temperature is somewhat higher, 
and our radial temperature profiles continue to rise at small radii
rather than flattening within $200$ kpc. 
Frenk et al.~(1999) note that this discrepancy is more or less present 
in all the grid-based codes but provide no explanation.  
To check whether this might be a result of numerical resolution or
algorithms employed, we recomputed the
Santa Barbara cluster with low, medium and high resolution initial
conditions corresponding to $64^3, 128^3$ and $256^3$ particles/cells
in the L0 grid, and L7 resolutions equal to 15.6, 7.8, and 3.9$h^{-1}$ kpc,
respectively. Instead of PPM, we also tested the artificial viscosity, 
finite-difference
algorithm for gas dynamics used in the ZEUS codes (Stone \& Norman 1992). 
In all cases we found the temperature
profile agreed with our result presented in Frenk et al. to within a 
few percent at all radii; i.e., our temperature profile has converged. 
Note also that there is excellent agreement between our results for the Santa Barbara
cluster and those of a new, completely independent AMR code (see Fig.1 of
Kravtsov, Klypin, \& Hoffman 2001).

\section{Results}

For each simulation, clusters were identified in the refined subvolume and 
their virial radii (r$_{vir}$) determined as the radius at which the mean 
enclosed density was $\Delta_c$ times the critical density (at $z=0$,
$\Delta_c \sim 101$ for $\Lambda$CDM and 178 for SCDM; e.g., Bryan \& Norman
1998). Projected gas density, X-ray surface brightness, and 
emission-weighted gas temperature images
(5 h$^{-1}$ Mpc on a side) were constructed for each cluster assuming
a Raymond-Smith spectrum with metallicity 0.3 solar in the (1.5-11.0) keV 
bandpass
used in the analysis of M98. 2D radial temperature profiles (centered on
the X-ray peak) were determined from the temperature images and a global
temperature (T$_x$) calculated by summing the emission-weighted temperature
within 1.0 h$^{-1}$Mpc. This procedure was intended to duplicate, 
as closely as possible, that of M98.

\begin{figure}[t]
\epsscale{0.85}
\plotone{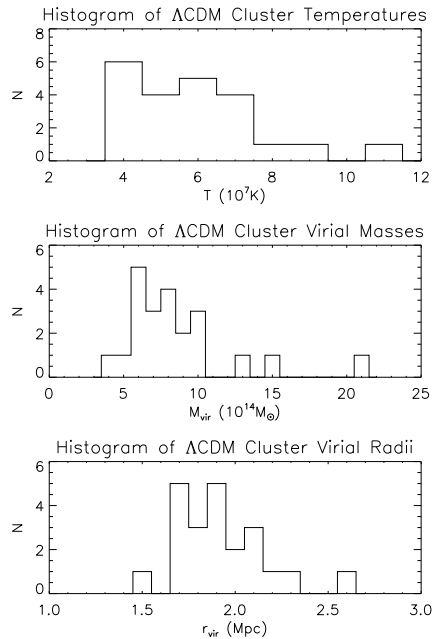}
\caption{
Histograms of the temperature, mass and virial radius distributions
for the $\Lambda$CDM sample of simulated clusters.
}
\label{hist}
\end{figure}

\begin{figure}[t]
\epsscale{0.92}
\plotone{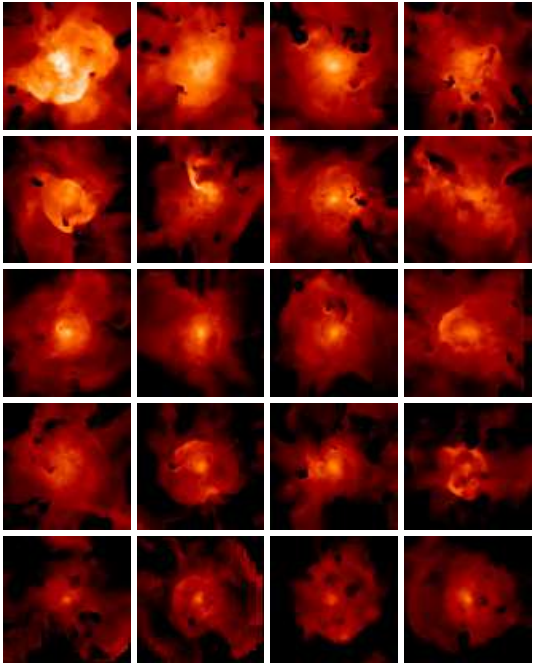}
%%\plotone{/scratch/cloken/AMR/PROFS/LCDM/temp_maps_80x80.ps}
\caption{Projected emission-weighted temperature maps of the
$\Lambda$CDM clusters at $z=0$. Each image is 
5 h$^{-1}$ Mpc on a side. Clusters are ordered from most massive (top left)
to least massive (bottom right).}
\label{tempmaps}
\end{figure}

The distribution of cluster temperatures (T$_x$), virial masses (M$_{vir}$),
and virial radii ($r_{vir}$) in the $\Lambda$CDM sample are shown in 
Fig.~\ref{hist}. The complex, non-isothermal structure of the clusters
is shown in the thumbnail images in 
Fig.~\ref{tempmaps}.
Although complicated, a
number of patterns can be discerned: the core of the cluster is generally
hotter than the outskirts; low-temperature ``holes" correspond to infalling
groups with lower masses and hence, lower temperatures.  In addition, a
few clusters show strong shocks due to recent major mergers, but these are
generally not spherically symmetric and tend to be of order a Mpc from the
core and so might be difficult to detect observationally. 
 The evolution of the temperature structure and the 
significant
differences between SCDM and $\Lambda$CDM clusters at $z=0.5$ will be 
discussed in a forthcoming paper (Loken et al. 2002).

\begin{figure}[h]
\epsscale{0.80}
\plotone{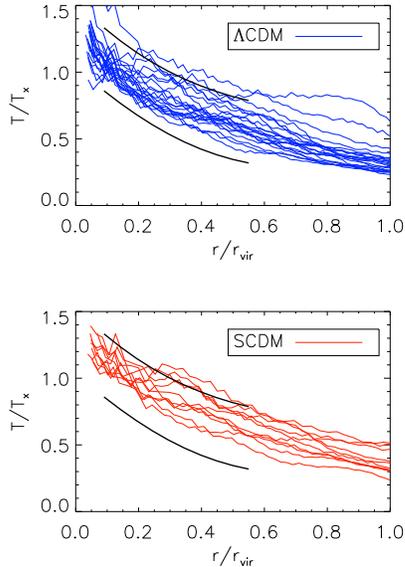}
\caption{
Temperature profiles for all the clusters in our $\Lambda$CDM (top)
and SCDM (bottom) samples. Temperatures and radii are scaled in terms of the
cluster's global temperature and virial radius, respectively. The heavy,
black lines are reproduced from Fig.~7 of M98 and enclose most of their
data points and error bars. }
\label{temp_all}
\end{figure}

\begin{figure}[h]
\epsscale{1.0}
\plotone{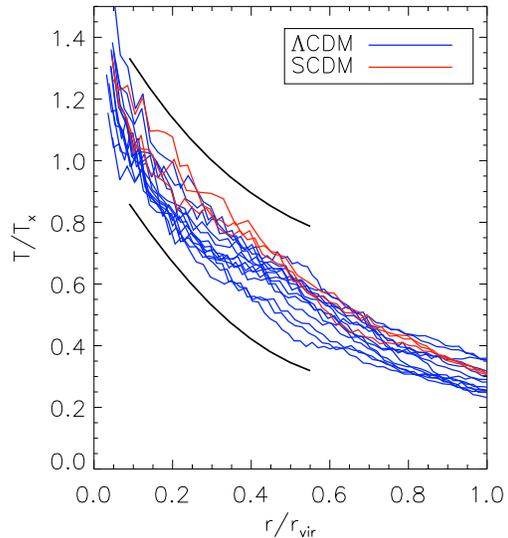}
\caption{
Temperature profiles for the symmetric clusters in our $\Lambda$CDM (blue)
and SCDM (red) samples. Scaling and the observational bounds from M98 are as in
Fig.~\ref{temp_all}.}
\label{profs}
\end{figure}

The key result of this current paper concerns the 2D temperature profiles
shown in Figs.~\ref{temp_all} and \ref{profs}. A given cluster's temperature profile is
normalized by its global temperature ($T_x$) and the radius is scaled
by the cluster's virial radius. Note that the resulting profiles 
are
remarkably similar even when comparing $\Lambda$CDM and SCDM clusters.
In addition, the majority of the profiles fall within the band drawn by M98 (reproduced
in Figs.~\ref{temp_all} and \ref{profs}) to  encompass
the majority of their data points and error bars.

Fig.~\ref{profs} is a``tidied'' version of Fig.~\ref{temp_all} which includes only those 
clusters whose X-ray
appearance is fairly regular and symmetric. This is exactly the same selection
cut imposed by M98 and others searching for a general temperature profile. 
However, even the excluded profiles are, for the most part, not dramatically
different in general appearance. The discarded profiles
generally have strong bumps or wiggles at the radius
corresponding to a merging subcluster. 13 of 22  $\Lambda$CDM cluster
profiles are kept in Fig.~\ref{profs} (similarly, M98 find that 17 of their 33
clusters are symmetric). However, only  3 of 10 SCDM cluster are sufficiently
regular; presumably this reflects the fact that clusters are still forming
at $z=0$ in an SCDM universe.

We find that the temperature profiles in Fig.~\ref{profs} can be 
well fit with an expression of the form:
$T=T_o [1+r/a_x ] ^{-\delta}$, where
$T_o=1.33$, $a_x = r_{vir}/1.5$, and 
$\delta = 1.6$ on the radial range $(0.04-1.0) \; r_{vir}$.
M98 chose instead to fit their temperature profiles with a polytropic equation 
of state:
$T=T_o [1+(r/a_x) ^2]^{-3 \beta (\gamma-1)/2}$.
Using this form on the same radial range as M98, $(0.09-0.55) \; r_{vir}$,
we find  
$T_o=1.15$, $a_x = r_{vir}/7.2$, and $\gamma=1.26$ (assuming
$\beta = 2/3$). M98 found a very similar fit: $\gamma=1.24^{+.20}_{-.12}$
and $a_x = r_{vir}/11$. We note, however, that the shape of the fit changes
when the fitted range is extended to $(0.04-1.0) \; r_{vir}$ in
which case: $T_o=1.17$, $a_x = r_{vir}/5$ and $\gamma=1.39$.
This sensitivity to the choice of fitting range appears to be due 
to the fact that the scatter in the data decreases for $r> \sim0.6 \; r_{vir}$
and suggests that observations must extend to near the virial radius
in order to converge on the true, universal temperature profile.

Though the two forms for the temperature profile both yield acceptable
fits, we prefer the first form for two reasons. First, the key difference 
between the two fits is that the polytropic form
turns over and begins to level-off at small radii while the first
one continues to rise (just as in Fig.~\ref{profs}). 
If the fits are extended within $r=0.09 \; r_{vir}$, then the polytropic
form yields an increasingly poor fit. It is difficult to draw any strong
conclusions, however, because this is exactly the region within which
non-adiabatic effects would become important and, as well, our limiting 
resolution is $\sim 0.01 \; r_{vir}$ for a typical cluster.

The second argument against the polytropic form is the fact that
it is not well-motivated physically. In fact, it can be derived by assuming
that the gas is: {\it isothermal}, follows a $\beta$-model for
the X-ray surface brightness, and obeys the {\it polytropic} equation of state,
$T=T_o \rho^{\gamma-1}$. Clearly, this is an inconsistent derivation.
Moreover, the assumption of a polytropic equation of state amounts
to assuming that the gas is isentropic which clearly contradicts the
outwardly rising entropy profiles which we and others find from
simulations (e.g., Frenk et al.~1999).  

\begin{figure}[t]
\epsscale{1.0}
\plotone{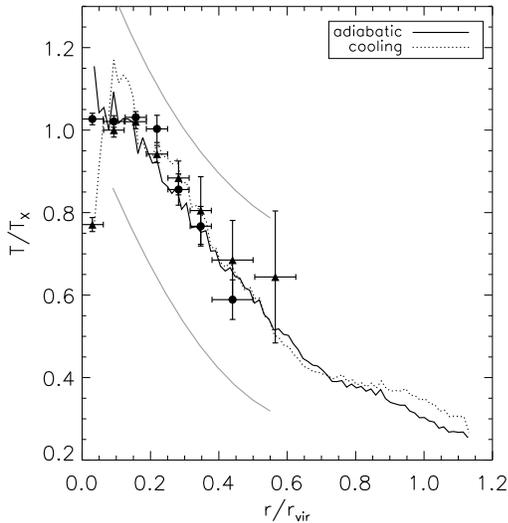}
\caption{Mean temperature profiles (with 1-$\sigma$ errors) for a sample of 11 cooling flow (filled triangles) and
10 non-cooling flow clusters (filled circles) observed by {\it Beppo}SAX (De Grandi \& Molendi 2002) along
with our numerical results for a particular cluster simulated with and without radiative cooling. The grey bands
represent the M98
observational bounds as in Fig.~\ref{temp_all}.}
\label{beppo}
\end{figure}

Recently, De Grandi \& Molendi (2002) have analyzed {\it Beppo}SAX data for a sample of 21 clusters
with and without cooling flows. They concluded that non-cooling flow
cluster temperature profiles are flat in the innermost regions 
($r < 0.2 \; r_{vir}$) and that the profiles of both types of clusters
decline rapidly with radius outside $r \sim 0.2 \; r_{vir}$. 
Their mean data points for the two
types of clusters are shown in Fig~\ref{beppo} along with profiles for two of our clusters which
are discussed in \S\ref{physics}. Again, the shape and normalization of the 
observational profiles agree well with the simulations though the flat innermost region
may signal the need for some additional physics in the simulations (e.g. De Grandi \& Molendi 2002).

\begin{figure}[t]
\epsscale{1.0}
\plotone{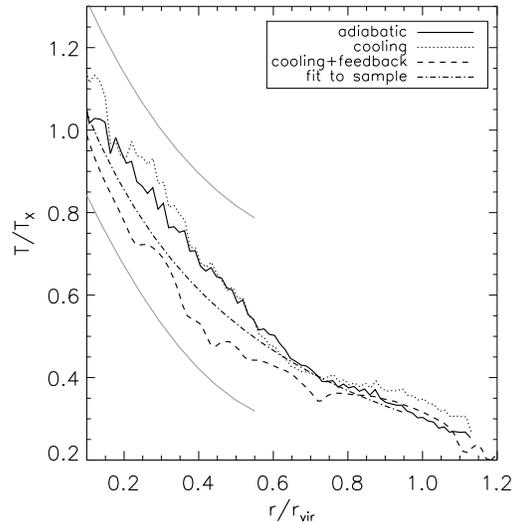}
\caption{
Temperature profiles for a cluster from our adiabatic sample (solid line) and the same
cluster rerun to include radiative cooling (dotted line). Our fit to the entire sample of
symmetric clusters is shown as the dash-dot line. Also shown is the ``Santa Barbara"
cluster from a simulation incorporating cooling and supernovae feedback. See text for more
details. The grey bands are again reproduced from Fig.~7 of M98 and enclose most of their
data points and error bars. }
\label{cooling}
\end{figure}

\subsection{Role of Additional Physics}
\label{physics}

The present adiabatic simulations are the first step in a planned series of simulations which
will incorporate non-adiabatic physical effects. Thus they form a baseline sample for 
evaluating the
significance of additional physics. Though potentially important processes like 
cooling and galaxy feedback are not included in the simulations for the adiabatic cluster catalog, 
we would expect any effects
to be strongly limited to the cluster core and to have very little
effect in the outer regions. These expectations now appear justified by our most recent
non-adiabatic simulations.

Two of the clusters in our sample have been re-run with identical parameters and resolution
in order to include the effects of radiative cooling (Motl et al. 2002). These new runs are
compared with the adiabatic simulations in Fig.~\ref{cooling}. The temperature profiles of
the cooling clusters are scaled as in the adiabatic case except that we omit the inner core 
($r<0.1\; r_{vir}$) when calculating the global temperature.  In addition, 
one of us (MLN) has recently performed a higher resolution (7.8$h^{-1}$kpc) simulation 
including both radiative cooling, and Type II supernovae feedback. This cluster, which
has the same initial parameters as the Santa Barbara comparison cluster (Frenk et al. 1999)
is also included on the figure for comparison. We caution that this is just a first
simulation and hence the full parameter space of star formation and feedback parameters has
not been explored.

Note that, as expected, the temperature profile of the cooling cluster is virtually indistinguishable 
from that of the adiabatic cluster outside of the cool core ($r>0.1\; r_{vir}$). 
This strongly suggests that our conclusions about declining cluster temperature profiles are
not the result of neglecting radiative cooling. Perhaps even more interesting is the 
higher-resolution simulation which incorporates both cooling and feedback. Again, the
shape of the corresponding temperature profile is remarkably similar to the other cases.
Fig.~\ref{cooling} makes clear that the non-adiabatic profiles also fit well within the
observational bounds determined by M98

Electron thermal conduction, which is not included in our simulations,
would have to be a significant fraction of its Spitzer value in
order to erase temperature gradients on the scale we observe.
The suppression of thermal conduction by tangled cluster magnetic fields
is an open area of research which has, as yet, reached no firm
conclusion (Chandran \& Cowley, 1998; Malyshkin \& Kulsrud, 2001).

\section{Discussion}

We have used our volume-limited catalogs of numerically-simulated clusters to
demonstrate that both $\Lambda$CDM and SCDM clusters follow a universal cluster 
temperature profile with  $T/T_o=1.3  [ 1 + 1.5 r/r_{vir} ] ^{-1.6}$. This temperature  profile
agrees very well with the observationally-determined profile of M98 and the more recent {\it Beppo}SAX data 
of DeGrandi \& Molendi (2002) (Fig.~\ref{beppo}). Our 
simulations also reveal a wealth of extremely complex and non-isothermal temperature structure
(Fig.~\ref{tempmaps}) which current X-ray telescopes may be able to probe.
Interestingly, our 
preliminary analysis suggests that the large-scale cluster temperature structure 
(particularly for $z>0$) is a potential discriminator of 
cosmological model.

Our current adiabatic catalogs of clusters were intended to serve as a well-defined baseline sample 
for statistical and individual comparison with both observed clusters and with future simulations
that include additional physical effects. 
Remarkably, our initial simulations of clusters with more realistic physics (radiative cooling
and galaxy feedback) give rise to temperature profiles (Fig.~\ref{cooling}) that agree well with our
adiabatic profile and suggest that this truly is a universal cluster temperature profile.

\noindent 
{\bf Acknowledgements}
 
This work has been partially supported by NASA grant NAG5-7404
and NSF grant AST-9803137. G.L.B. is supported by NASA through
Hubble Fellowship grant HF-01104.01-98A from the Space Telescope
Science Institute, which is operated under NASA contract NAS6-26555.
The simulations were carried out on
the SGI/Cray Origin 2000 at the National Center for Supercomputing
Application, University of Illinois Urbana-Champaign.

\clearpage

\end{document}